\documentclass[reprint,twocolumn,aps,prc,a4paper,superscriptaddress,showpacs,preprintnumbers,amsmath,amssymb]{revtex4}
\usepackage{color}
\usepackage{graphicx}
\usepackage{CJK}
\usepackage{longtable,booktabs}
\usepackage{ltablex}
\begin{document}

\title{Mixing of one-particle-one-hole projected states with the variation after projection wave functions}

\author{Xiao Lu}
\affiliation{China Institute of Atomic Energy, Beijing 102413, China}
\author{ Zhan-Jiang Lian}
\affiliation{China Institute of Atomic Energy, Beijing 102413, China}
\author{ Zao-Chun Gao}
\affiliation{China Institute of Atomic Energy, Beijing 102413, China}

\begin{abstract}
 In this paper, we study the mixing of one-particle-one-hole projected states with the variation after projection (VAP) wave functions in attempt to improve the approximation of this method. It turns out that when minimizing only the lowest (yrast) energy with given spin and parity, the one-particle-one-hole projected states can not be mixed into the converged VAP wave function, which is very similar to the situation of the Hartree-Fock method. However, if one minimizes the sum of several lowest energies with the same spin and parity, the one-particle-one-hole mixing can make some improvements to the VAP wave functions. We expect such one-particle-one-hole mixing may be useful in the calculations of the low-lying excited states in heavy nuclei with a large model space.

\end{abstract}

\maketitle
\section{Introduction}
The nuclear shell model (SM) \cite{Caurier05} has been very successful in describing various properties of nuclei, especially in the neighborhood of the closed shells, where the configuration space is usually small but good enough for the construction of the nuclear wave function. However, in heavy deformed nuclear region, the configuration space is huge, in which the full SM calculation is almost impossible. Such huge configuration space must be compressed so that SM calculation can be performed on a present-day computer. Unfortunately, the energies and wave functions 
obtained in a compressed configuration space are approximated ones. To make such approximated solutions as close as possible to the exact shell model ones, various methods, such as shell model truncation \cite{ST94}, stochastic quantum Monte Carlo approaches \cite{MC97,MC01}, projected configuration interaction \cite{PCI09}, and the class of variation after projection (VAP) methods \cite{VAMPIR04,Shimizu21,VAP15,VAP17,VAP18,VAP21}, have been developed.

Among those approximated SM methods, the VAP method is an important one, in which the nuclear wave functions with good quantum numbers are sufficiently optimized \cite{VAMPIR04,Shimizu21,VAP15,VAP17,VAP18,VAP21}. In Refs. \cite{VAMPIR04,Shimizu21}, the Hartree-Fock-Bogoliubov (HFB) vacuum states are adopted, and the neutron and proton number projections are performed in addition to the angular momentum projection. Very good approximations of their results have been achieved. However, the full projection of the HFB state onto good quantum numbers (i.e., neutron number, proton number, angular momentum and parity) requires five-fold integration, which is too much time consuming. On the other hand, in Ref. \cite{VAP15}, we have shown that the angular momentum projection is crucial in obtaining good shell model approximation. So, a simpler VAP can be the one in which the Hartree-Fock (HF) Slater determinants (SDs) are taken and only the angular momentum projection and parity projection are performed \cite{VAP17,VAP18,VAP21}. Apparently, with the same number of projected states, the approximation of the VAP with projected SDs is not as good as that with projected HFB vacua. This is because HFB vacuum incorporates more correlations especially pairing. However, the approximation of the former can be continuously improved by adding more and more projected SDs. In this sense, we are more interested in the VAP with projected SDs due to its relatively less computational cost. Thus we only discuss the VAP with projected SDs in this paper.

In Ref. \cite{VAP18}, we have proposed a new algorithm of the VAP calculation, in which the yrast state and non-yrast states can be varied on the same footing. In this algorithm, the orthogonality among the calculated states is automatically fulfilled by solving the Hill-Wheeler (HW) equation. This avoids the complexity of the frequently used Gram-Schmidt orthogonalization, as adopted in Ref. \cite{VAMPIR04}. After that, we further found that the complicated $K$-mixing can be safely removed from VAP with the adopted SDs fully symmetry-unrestricted. This considerably simplifies the VAP calculation, especially in the application to the high-spin states with arbitrary deformation \cite{VAP21}.

However, if the VAP method is applied to heavy deformed nuclei, the computational burden becomes quite heavy. Under this situation, the VAP calculation with large number of projected states becomes impractical. A further improvement of such VAP wave function may be the inclusion of the projected particle-hole states built on top of the converged reference SDs. This is very similar to the cases of the Projected Shell Model \cite{SY95} and the MONSTER method \cite{VAMPIR04}, in which the projeced quasi-particle states on top of the selected projected vacuum are included to form the nuclear wave function.

The mixing of particle-hole projected states with the VAP wave functions requires more complicated projected matrix elements to form a wider Hill-Wheeler (HW) equation. Fortunately, in the present VAP calculation \cite{VAP17,VAP18,VAP21}, all the matrix elements needed in the one-particle-one-hole (1p-1h) mixing are already available. These matrix elements are originally used to build the Hessian matrix of the VAP energy, so that the VAP calculation may converge after a very few iterations (see Fig. \ref{fig1} below). Thus the improved energies and the nuclear wave functions with 1p-1h mixing can be calculated without much computational cost. However, if one further considers the 2p-2h mixing, there are two troubles. The first one is,  new projected matrix elements among the 2p-2h configurations must be calculated specifically. The other one is the number of the 2p-2h SDs is much larger than the 1p-1h one. Therefore the calculation including 2p-2h mixing is much more complicated. Here, for simplicity, we only discuss the 1p-1h mixing in this paper and see how it affects the VAP wave function.

The paper is organized as follows. Section \ref{vap} provides a general introduction of the adopted method. Section \ref{yrast} discusses the 1p-1h mixing after the VAP for the yrast states. Section \ref{nonyrast} discusses the 1p-1h mixing after the VAP for the non-yrast states. Summary and outlook are presented in Section \ref{sum}.

\section{The VAP Method and the 1p-1h mixing}\label{vap}

Let's start the introduction of the present method with the simplest case that only a single SD, $|\Phi\rangle$, is considered. At given quantum numbers of spin, $J$, parity, $\pi$, and magnetic quantum number, $M$, one can generate $2J+1$ projected states, $P_{M K}^{J\pi}|\Phi\rangle$, from $|\Phi\rangle$. Here, $P_{M K}^{J\pi}$ stands for the product of the angular momentum projection operator, $P_{M K}^{J}$, and the parity projection operator, $P^\pi$. Generally, all these $2J+1$ projected states are expected to be taken to form the trial nuclear wave function,
\begin{equation}\label{wf}
|\Psi_{J\pi M\alpha}\rangle=\sum_{K=-J}^{J} f_{K}^{J\pi\alpha} P_{M K}^{J\pi}|\Phi\rangle,
\end{equation}
where, $\alpha$ is used to differ the states with the same $J$, $\pi$ and $M$.  The coefficients, $f_{K}^{J\pi\alpha}$, and the corresponding energy, $E^{J\pi}_\alpha$, are determined by solving the following Hill-Wheeler (HW) equation,
\begin{eqnarray}\label{hw0}
\sum_{K'=-J}^{J}(H^{J\pi}_{KK'}-E^{J\pi}_\alpha N^{J\pi}_{KK'})f^{J\pi\alpha}_{K'}=0,
\end{eqnarray}
where $H^{J\pi}_{KK'}=\langle\Phi|\hat H P^{J\pi}_{KK'}|\Phi\rangle$ and $N^{J\pi}_{KK'}=\langle\Phi|P^{J\pi}_{KK'}|\Phi\rangle$. For convenience, we assume $E^{J\pi}_1\leq E^{J\pi}_2\leq\cdots\leq E^{J\pi}_{2J+1}$. The coefficients $f^{J\pi\alpha}_{K}$ should satisfy the normalization condition,
\begin{eqnarray}\label{norm0}
\sum_{K,K'=-J}^{J}f^{J\pi\alpha*}_{K}N^{J\pi}_{KK'}f^{J\pi\alpha}_{K'}=1.
\end{eqnarray}

 $|\Psi_{J\pi M\alpha}\rangle$ needs to be optimized so that it can be as close as possible to the exact shell model one. According to Eq. (\ref{hw0}), $|\Psi_{J\pi M\alpha}\rangle$ is completely determined by $| \Phi\rangle$ provided the Hamiltonian is given. Thus the optimization of Eq. (\ref{wf}) is actually realized by varying the $|\Phi\rangle$ state.

 To vary the $|\Phi\rangle$ state, one may first need to parameterize it. Suppose there is a normalized HFB vacuum state, $|\Phi_0\rangle$. The corresponding quasiparticle operators are denoted by
$\beta^\dagger_{0,\mu}$ and $\beta_{0,\mu}$. Using the Thouless theorem \cite{TH60,PR80}, $|\Phi_0\rangle$ can be changed to a new HFB vacuum state $|\Phi\rangle$, namely,
\begin{equation}\label{thls}
|\Phi\rangle=\mathcal{N} e^{\frac{1}{2} \sum_{\mu \nu} d_{\mu \nu} \beta_{0,\mu }^{\dagger}\beta_{0,\nu }^{\dagger}}|\Phi_{0}\rangle=
\mathcal{N} e^{\sum_{\kappa} d_{\kappa} A^{\dagger}_{\kappa }}|\Phi_{0}\rangle,
\end{equation}
where $d$ is a complex skew matrix. The matrix elements $d_{\mu \nu}$ will be considered as the variational parameters. $\mathcal{N}$ is the normalization parameter, so that $\langle\Phi|\Phi\rangle=1$. For convenience, the subscript $\kappa$ is used to stand for the $(\mu,\nu)$ numbers with $\mu<\nu$, and the particle pair operators are defined as,
\begin{eqnarray}
 \hat{A}_{\kappa}^{\dagger}&=&\beta_{0,\mu}^{\dagger}\beta_{0,\nu}^{\dagger},\\
 \hat{A}_{\kappa}&=&(\beta_{0,\mu}^{\dagger}\beta_{0,\nu}^{\dagger})^\dagger=\beta_{0,\nu}\beta_{0,\mu}.
\end{eqnarray}
Here, the vacuum states $|\Phi_0\rangle$ and $|\Phi\rangle$ are reduced to be SDs, so that the particle number projections can be omitted. Therefore, the $\hat{A}_{\kappa}^{\dagger}$ operators  become 1p-1h operators corresponding to the $|\Phi_{0}\rangle$ state.

Now, let's present a brief introduction on how to establish the VAP iteration. Suppose that we have a $|\Phi\rangle$ state at certain VAP iteration. Then we set $|\Phi_0\rangle=|\Phi\rangle$ and $d=0$. At this $d=0$ point, the energy, $E^{J\pi}_\alpha$, and the corresponding $f_{K}^{J\pi\alpha}$ coefficients are calculated first by solving Eq. (\ref{hw0}). This process includes two successive diagonalizations. The first one is the diagonalization of the norm matrix $N^{J\pi}$ in Eq. (\ref{hw0}) and we have
\begin{eqnarray}\label{dn}
\sum_{K'=-J}^J N^{J\pi}_{KK'} R^k_{K'}=\sigma_kR^k_K,
\end{eqnarray}
 where $\sigma_k  \geq 0$ and $R^k$ with $k=1,2,...2J+1$ are eigenvalues and the corresponding eigenvectors, respectively. Then one can establish a new set of orthonormal basis states, $|\psi_k^{J\pi}\rangle \,(k=1,2,\cdots,2J+1)$,
 \begin{eqnarray}\label{psi}
 |\psi_k^{J\pi}\rangle=\frac {1}{\sqrt{\sigma_k}} \sum_{K=-J}^J R^k_K P^{J\pi}_{MK}|\Phi\rangle,
 \end{eqnarray}
and Eq. (\ref{hw0}) can be transformed into a normal eigenvalue equation
\begin{eqnarray}\label{eigen}
\sum_{k'=1}^{2J+1}\left [\langle \psi_k^{J\pi}|\hat H|\psi_{k'}^{J\pi}\rangle-E^{J\pi}_\alpha \delta_{kk'}\right ]u^{J\pi\alpha}_{k'}=0.
\end{eqnarray}
The second diagonalization is then performed on Eq. (\ref{eigen}), so that  the energies, $E^{J\pi}_\alpha$, in Eq. (\ref{hw0}) are obtained and the coefficients of the wave functions, $f^{J\pi\alpha}_{K}$, are transformed from $u^{J\pi\alpha}_{k}$, i.e.,
\begin{eqnarray}\label{eigen1}
f^{J\pi\alpha}_{K}=\sum_{k=1}^{2J+1} \frac{R^k_K u^{J\alpha}_k}{\sqrt{\sigma_k}}.
\end{eqnarray}

 Notice that, in Eq. (\ref{psi}), $\sigma_k$ should not be zero, or the corresponding $|\psi_k^{J\pi}\rangle$ should be abandoned. However, if $\sigma_k$ is too small,  the $|\psi_k^{J\pi}\rangle$ basis states may not be calculated precisely enough to guarantee the stability of the calculated energies and wave functions. This is one of the main troubles when all $2J+1$ projected basis states are included in the VAP calculation, because some of the smallest $\sigma_k$ values could become tiny and tiny and finally the VAP iteration collapses.

After we have the projected energies, $E^{J\pi}_\alpha$, and the corresponding wave functions, we chose the lowest one, $E^{J\pi}_1$, and minimize it. Thus the gradient and the Hessian of $E^{J\pi}_1$ in the space spanned by the $d_{\kappa}$ parameters are calculated according to the formulas in Ref. \cite{VAP17}, then the next improved SD $|\Phi\rangle$ can be determined by adopting the trust region Newton's algorithm \cite{numopt}. The VAP iteration terminates when the gradient of $E^{J\pi}_1$ becomes less than 0.01keV, which is precise enough to obtain the exact minimum of $E^{J\pi}_1$.

\begin{figure}[ht!]
 \centering
 \includegraphics[width=8cm]{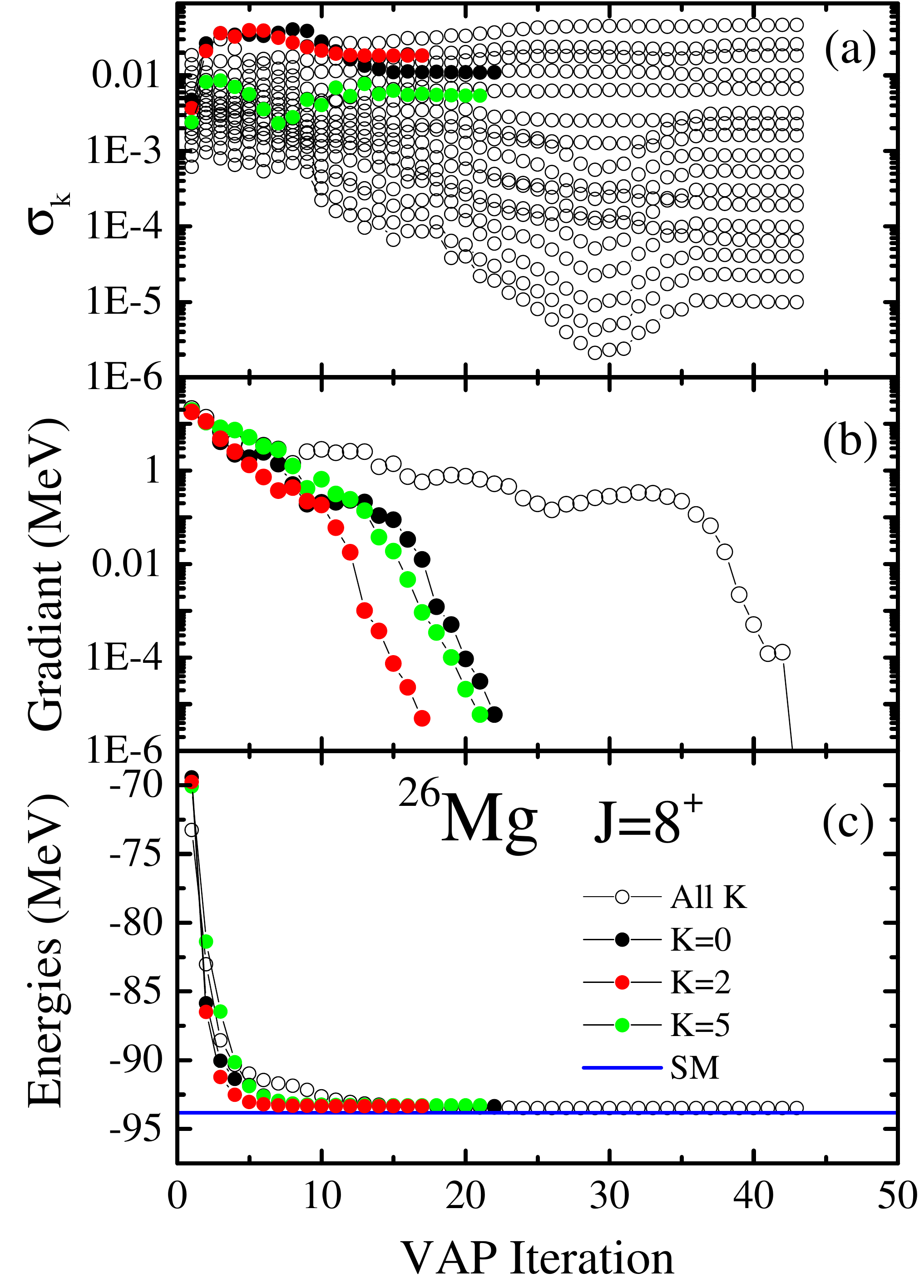}
 \caption{\label{fig1}(Color online) Quantities as functions of the VAP iteration for the $J^\pi$= 8$^{+}$ state in $^{26}$Mg. Results with Eq. (\ref{wf}) are shown as open circles. Those with Eq. (\ref{wfk}) are shown as filled dots in different colors. The calculated quantities are (a) the norms, $\sigma_k$; (b)the absolute values of the energy gradients; and (c) the VAP energies. The iteration terminates once the absolute value of the  energy gradient is less than 0.01keV. The USDB interaction is adopted.
}
\end{figure}

In Fig. \ref{fig1}, we demonstrate such VAP iteration using the example of the yrast $8^+$ state in $^{26}$Mg, which will be further discussed in the next Section. The initial SD $|\Phi\rangle$ is randomly chosen.  The adopted Hamiltonian is USDB \cite{usdb}. The results are shown as black circles. From this initial $|\Phi\rangle$, the VAP converges normally after 43 iterations.

However, according to Ref. \cite{VAP21}, the form of the trial VAP wave function with spin, $J$, can be simplified by adopting just one projected state rather than adopting all $2J+1$ angular momentum projected states for each selected reference state. To confirm this point, we chose only one of the $2J+1$ projected states in Eq. (\ref{wf}), and the trial VAP wave function can be considerably simplified as,
\begin{equation}\label{wfk}
|\Psi_{J\pi M} (K)\rangle= \frac{P_{MK}^{J\pi}|\Phi\rangle}{\sqrt{\langle\Phi|P_{KK}^{J\pi}|\Phi\rangle}},
\end{equation}
where, the $K$ number can be randomly chosen. Here, we still take the above example and choose $K=0,2$ and $5$ to optimize Eq. (\ref{wfk}) independently by using exactly the same VAP code. The results are shown as filled dots with different colors in Fig. \ref{fig1}. It is shown that all the converged energies for $|\Psi_{J\pi M} (K)\rangle$ with different $K$ are almost the same and also very close to the converged $E^{J\pi}_1$.

 Now, we have one optimized wave function of Eq. (\ref{wf}) and three optimized ones of Eq. (\ref{wfk}) with $K=0,2$ and $5$, which must be compared with one another. Overlaps among these wave functions are calculated. The values of $|\langle\Psi_{J\pi M1}|\Psi_{J\pi M} (K)\rangle|$ corresponding to the converged results in Fig. \ref{fig1} are 0.9867,0.9853,0.9898 for $K=0,2$ and $5$, respectively.
Actually, results with other $K$ numbers are almost the same as in Fig. \ref{fig1}. This clearly shows the equivalence of Eq. (\ref{wf}) and Eq. (\ref{wfk}) and implies the equivalence among the $|\Psi_{J\pi M} (K)\rangle$ states themselves. The latter is confirmed by calculating the quantities of $|\langle\Psi_{J\pi M}(K)|\Psi_{J\pi M} (K')\rangle|$, which turn out to be  0.9930, 0.9911 and 0.9875 for $(K,K')=$(0,2), (0,5) and (2,5), respectively.

We should stress that the above equivalences are valid only for the optimized VAP wave functions. With the same $|\Phi\rangle$, Eq. (\ref{wf}) is expected to be better than Eq. (\ref{wfk}). This can be seen in Fig. \ref{fig1} that the energy of Eq. (\ref{wf}) is indeed much lower than the ones of Eq. (\ref{wfk}) at the first iteration where the same initial  $|\Phi\rangle$ is used to start all the VAP calculations. However, in the process of VAP iteration, the $|\Phi\rangle$ state in $|\Psi_{J\pi M1}\rangle$ and those ones in the $|\Psi_{J\pi M} (K)\rangle$  states change independently, and they may become different from one another. This clearly tells us the converged VAP wave functions with different reference SD states can be almost equivalent.

The advantages of introducing Eq. (\ref{wfk}) into VAP are obvious. First of all, the number of all required matrix elements with Eq. (\ref{wfk}) is only one $(2J+1)^2$th of that with Eq. (\ref{wf}). Consequently, the computational time with Eq. (\ref{wfk}) is remarkably reduced. In the example of Fig. \ref{fig1}, the elapsed time for each iteration with Eq. (\ref{wfk}) is only about 3s on an Intel Xeon CPU with 20 cores, while such time with Eq. (\ref{wf}) is about 360s on the same CPU. Secondly, the VAP iteration with Eq. (\ref{wfk}) converges more reliably and faster than that with Eq. (\ref{wf}). As one can see from Fig. \ref{fig1}(a), the smallest $\sigma_k$ values corresponding to Eq. (\ref{wf}) tend to be tiny and tiny as the iteration proceeds. Fortunately, they come back after the 30th iteration in this calculation. However, this is not always lucky. In many cases, the $\sigma_k$ values could keep on moving to zero and one can not get a converged VAP wave function with Eq. (\ref{wf}) (See the examples in the supplemental material for Ref. \cite{VAP21}). Finally, the VAP calculation with Eq. (\ref{wfk}) can be conveniently extended to arbitrary high-spin states, while the VAP with Eq. (\ref{wf}) is very difficult to do that.

Recognizing the equivalence of Eq. (\ref{wf}) and Eq. (\ref{wfk}) in VAP, we prefer to take the simplified Eq. (\ref{wfk}) and extend it by including more reference SDs,
\begin{equation}\label{wfks}
|\Psi_{J\pi M \alpha}(K)\rangle=\sum_{i=1}^{n} f_{i}^{J\pi\alpha} P_{M K}^{J\pi}|\Phi_{i}\rangle.
\end{equation}
 
The coefficients $f_{i}^{J\pi\alpha}$ and the corresponding energy, $E^{J\pi}_\alpha$, can be obtained by solving the following HW equation,
\begin{equation}\label{hw}
\sum_{i'=1}^{n}\langle\Phi_{i}|(\hat{H}-E^{J\pi}_\alpha) P_{K K}^{J\pi}| \Phi_{i'}\rangle f_{i^{\prime}}^{J\pi\alpha}=0.
\end{equation}
These $f_{i}^{J\pi\alpha}$ coefficients also should satisfy the normalization condition,
\begin{equation}\label{norm}
\sum_{ii'=1}^{n} f_{i}^{J\pi\alpha*}\langle\Phi_{i}| P_{KK}^{J\pi}| \Phi_{i^{\prime}}\rangle f_{i'}^{J\pi\alpha}=1.
\end{equation}

Here and below, all $E^{J\pi}_\alpha$ energies are calculated from Eq. (\ref{hw}) rather than Eq. (\ref{hw0}) since we no longer use Eq. (\ref{wf}). In this paper, we assume $E^{J\pi}_1\leq E^{J\pi}_2\leq\cdots \leq E^{J\pi}_n$. But in practical calculations we are only interested in the lowest $m (\leq n)$ energies.

From Eq. (\ref{hw}), one can see that changes of the $|\Phi_{i}\rangle$ states may directly change the $E^{J\pi}_\alpha$ energies. Our target is to try to find a set of these $n$ $|\Phi_{i}\rangle$ states, so that the calculated lowest $m$ $E^{J\pi}_\alpha$ energies can be as close as possible to the corresponding exact shell model ones denoted by $e_\alpha^{J\pi}$. Here, we also assume $e^{J\pi}_1\leq e^{J\pi}_2\leq\cdots \leq e^{J\pi}_m$.

Thanks to the Cauchy's interlacing theorem, which makes it clear that $E_\alpha^{J\pi}\geq e_\alpha^{J\pi}$ for any excited energies. This has been strictly proved in Ref. \cite{VAP18}. Recently, we just recognized there is a famous Hylleraas-Undheim-MacDonald (HUM) theorem widely known in the field of quantum chemistry \cite{hum1,hum2}. This theorem exactly tells us the relation of $E_\alpha^{J\pi}\geq e_\alpha^{J\pi}$.

Therefore, one can define the non-negative energy difference $\delta E_\alpha=E_\alpha^{J\pi}-e_\alpha^{J\pi}$. Then the summation of $\delta E_\alpha$ for  the lowest $m$ states must be non-negative, too. Namely,
\begin{eqnarray}\label{dem}
\Delta E_m=\sum_{\alpha=1}^m\delta E_\alpha=\sum_{\alpha=1}^m E_\alpha^{J\pi}-\sum_{\alpha=1}^m e_\alpha^{J\pi} \geq 0.
\end{eqnarray}
Clearly, if $\Delta E_m=0$, then $\delta E_\alpha=0$ for all included states. This means  $E_\alpha^{J\pi}=e_\alpha^{J\pi}$ and we obtain the exact eigenenergies of the given $\hat H$. But in general, $\Delta E_m >0$. Then our target is to try to find a number ($n$) of basis states through variation, so that $\Delta E_m$ becomes a minimum. Then the corresponding $E_\alpha^{J\pi}$ energies at the minimum of $\Delta E_m$ can be simultaneously obtained. These $E_\alpha^{J\pi}$ energies can be compared with the exact $e_\alpha^{J\pi}$ ones to test the quality of the present VAP method \cite{VAP18}. When $m=1$, we come back to the Ritz variational principle.

However, $\Delta E_m$ is usually unknown because we don't know the exact $e_i^{J\pi}$ energies in most cases. Actually, with a given $\hat H$, the $e_i^{J\pi}$ energies are determined, and can be considered as constants in Eq. (\ref{dem}). So minimizing $\Delta E_m$ is equivalent to minimizing
\begin{eqnarray}
S_m=\sum_{\alpha=1}^m E_\alpha^{J\pi}.
\end{eqnarray}

Since the involved $|\Phi_{i}\rangle$ states are varied independently, each of them has its own $d$ matrix, which is denoted by $d_i$. The matrix elements of $d_i$ are then denoted by $d_{i\kappa}$, namely

\begin{equation}\label{thlsi}
|\Phi_i\rangle=\mathcal{N}_i e^{\sum_{\kappa} d_{i\kappa} A^{\dagger}_{i\kappa }}|\Phi_{0,i}\rangle.
\end{equation}

For simplicity, we use $\vec{d}$ to denote the vector of all the independent $d_{i\kappa}$
parameters. Given a certain $\vec{d}$, all $|\Phi_{i}\rangle$ states can be uniquely determined. Consequently, the quantity $S_m$, determined by the set of $|\Phi_{i}\rangle$ states,  can be considered as a function of $\vec{d}$. Therefore, the VAP calculation is actually the minimization of $S_m$ in the space spanned by all the $d_{i\kappa}$ parameters.

Let us give a brief introduction to the present VAP iteration. Starting from a set of randomly chosen $|\Phi_{0,i}\rangle$ states, the gradient and the Hessian matrix of $S_m$ at $\vec{d}=0$ are calculated. Then a new point $\vec{d}_{\text min} \neq 0$ can be searched so that $S_m$ is as low as possible. This $\vec{d}_{\text min}$ determines a new set of $|\Phi_{i}\rangle$ states. We then update the $|\Phi_{0,i}\rangle$ states with these $|\Phi_{i}\rangle$ ones and perform the next VAP iteration. Notice that we only calculate the gradient and the Hessian matrix at $\vec{d}=0$. This is because at this special point, $|\Phi_{i}\rangle=|\Phi_{0,i}\rangle$, and the particle-hole operators $\hat{A}_{i\kappa}^{\dagger}$ become the ones corresponding to $|\Phi_{i}\rangle$. This may considerably simplify the VAP formulation and the corresponding calculation.

Now, let us consider the 1p-1h mixing. Suppose that we have a set of reference SD states, $|\Phi_{i}\rangle$, then the expanded wave function including 1p-1h components can be written as,
\begin{eqnarray}\label{wf1p1h}
|\Psi_{J\pi M \alpha}^{\prime}(K)\rangle&=&\sum_{i=1}^{n} {f'}_{i}^{J\pi \alpha} P_{M K}^{J\pi}|\Phi_{i}\rangle\nonumber\\
&&+\sum_{i=1}^{n} \sum_{\kappa} {f'}_{i\kappa}^{J\pi \alpha} P_{M K}^{J\pi} A_{i\kappa }^{\dagger}|\Phi_{i}\rangle.
\end{eqnarray}
Here, the $A_{i\kappa }^{\dagger}$ operators are the 1p-1h operators corresponding to $|\Phi_{i}\rangle$.  The coefficients, ${f'}_{i'(\kappa')}^{J\pi \alpha}$, and the corresponding energy, ${E'}^{J\pi}_\alpha$, can be obtained by solving the following expanded HW equations,
\begin{equation}\label{hwexp}
\left\{
\begin{aligned}[l]
&\sum_{i'=1}^{n}
 \langle\Phi_{i}|(\hat{H}-{E'}^{J\pi}_\alpha) P_{K K}^{J\pi}| \Phi_{i'}\rangle{f'}_{i'}^{J\pi \alpha} \\
 &+\sum_{i'=1}^{n} \sum_{\kappa'} \langle\Phi_{i} |(\hat{H}-{E'}^{J\pi}_\alpha) P_{K K}^{J\pi} A_{i'\kappa' }^{\dagger} |\Phi_{i'} \rangle{f'}_{i'\kappa'}^{J\pi\alpha}
  =0, \\
&\sum_{i'=1}^{n}
\langle\Phi_{i}| A_{i\kappa} P_{K K}^{J\pi}(\hat{H}-{E'}^{J\pi}_\alpha)|\Phi_{i'}\rangle {f'}_{i'}^{J\pi \alpha} \\
&+\sum_{i'=1}^{n} \sum_{\kappa'}\langle\Phi_{i}| A_{i\kappa }(\hat{H}-{E'}^{J\pi}_\alpha)P_{K K}^{J\pi} A_{i'\kappa' }^{\dagger}|\Phi_{i'}\rangle{f'}_{i'\kappa'}^{J\pi \alpha}
  =0. \\
\end{aligned}
\right.
\end{equation}

The coefficients, ${f'}_{i'(\kappa')}^{J\pi \alpha}$, should also satisfy the normalization condition, so that $ \langle\Psi_{J\pi M \alpha}^{\prime}(K) |\Psi_{J\pi M \alpha}^{\prime}(K) \rangle=1$.  Again, we assume $E'^{J\pi}_1\leq E'^{J\pi}_2\leq\cdots \leq E'^{J\pi}_m$.
Notice that all the matrix elements appearing in Eq. (\ref{hwexp}) are already available in our VAP calculations, because they are needed in the evaluation of the gradient and the Hessian matrix of $S_m$.

\section{1p-1h mixing with the yrast states}\label{yrast}
Let us first study the simplest case of Eq. (\ref{wfk}). The energy corresponding to Eq. (\ref{wfk}) can be expressed as
\begin{equation}\label{epj}
E=\frac{\langle\Phi|  \hat{H}P^{J\pi}_{KK}|\Phi\rangle}{\langle\Phi|P^{J\pi}_{KK}|\Phi\rangle}.
\end{equation}
Clearly, $E$ is a functional of $|\Phi\rangle$ governed by the complex $d_\kappa$ parameters through Eq. (\ref{thls}). Here, $d_\kappa$ can be explicitly written as
\begin{equation}\label{dk}
d_{\kappa}=x_{\kappa}+i y_{\kappa},
\end{equation}
where, $x_{\kappa}$ and $y_{\kappa}$ are real.
We assume $E$ reaches a minimum at $d=0$, then the gradient of $E$ at this point should be zero. Namely,
\begin{eqnarray}\label{px1}
&&\left.\frac{\partial E}{\partial x_{\kappa}}\right|_{\vec{d}=0}\nonumber\\
 &=&\left\{\langle\Phi|A_{\kappa}(\hat H-E)P^{J\pi}_{KK}|\Phi\rangle\right. \nonumber\\ &&\left.+\langle\Phi|(\hat{H}-E)P^{J\pi}_{KK}A^\dagger_{\kappa}|\Phi\rangle\right\} \frac{1}{\langle\Phi|P^{J\pi}_{KK}|\Phi\rangle}\nonumber\\
&=&0;\\
\label{py1}
&&\left.\frac{\partial E}{\partial y_{\kappa}}\right|_{\vec{d}=0}\nonumber\\
 &=&i\left\{-\langle\Phi|A_{\kappa}(\hat H-E)P^{J\pi}_{KK}|\Phi\rangle\right. \nonumber\\ &&\left.+\langle\Phi|(\hat{H}-E)P^{J\pi}_{KK}A^\dagger_{\kappa}|\Phi\rangle\right\} \frac{1}{\langle\Phi|P^{J\pi}_{KK}|\Phi\rangle}\nonumber\\
&=&0.
\end{eqnarray}
This immediately leads to the following equation
\begin{eqnarray}\label{grad}
\langle\Phi|(\hat{H}-E)P^{J\pi}_{KK}A^\dagger_{\kappa}|\Phi\rangle=0.
\end{eqnarray}

On the other side, it is known that in the Hartree-Fock theory, if the HF energy
\begin{equation}\label{ehf}
E_{HF}=\frac{\langle\Phi|  \hat{H}|\Phi\rangle}{\langle\Phi|\Phi\rangle}
\end{equation}
reaches a minimum, then one can easily see that no 1p-1h state on top of $|\Phi\rangle$ can be mixed into $|\Phi\rangle$, which means
\begin{equation}\label{hfcp}
\langle\Phi|\hat{H}A^\dagger_{\kappa}|\Phi\rangle=0.
\end{equation}
This inspired us that similar conclusion in the present VAP could be true. But here the projection operator is involved. The projected 1p-1h state
\begin{eqnarray}
|\Psi_{\kappa}\rangle\equiv P^{J\pi}_{MK}A^\dagger_{\kappa}|\Phi\rangle
\end{eqnarray}
 is no longer orthogonal to $|\Psi_{J\pi M}(K)\rangle$. One can do the Gram-Schmid orthogonalization to get a new state
\begin{eqnarray}
|\Psi'_{\kappa}\rangle=|\Psi_{\kappa}\rangle
-\langle \Psi_{J\pi M}(K)|\Psi_{\kappa}\rangle|\Psi_{J\pi M}(K)\rangle,
\end{eqnarray}
so that
\begin{eqnarray}
\langle \Psi_{J\pi M}(K)|\Psi'_{\kappa}\rangle=0.
\end{eqnarray}
If there is no mixing of the 1p-1h projected state $|\Psi_{\kappa}\rangle$ with $|\Psi_{J\pi M}(K)\rangle$, then one should expect that
\begin{eqnarray}\label{mx0}
\langle \Psi_{J\pi M}(K)|\hat H|\Psi'_{\kappa}\rangle=0.
\end{eqnarray}
It is easy to prove that Eq. (\ref{mx0}) is exactly equivalent to Eq. (\ref{grad}). This clearly tells us that the property of no 1p-1h mixing at an energy minimum can be generalized from HF theory to VAP.

\begin{figure}
 \centering
 \includegraphics[width=8cm]{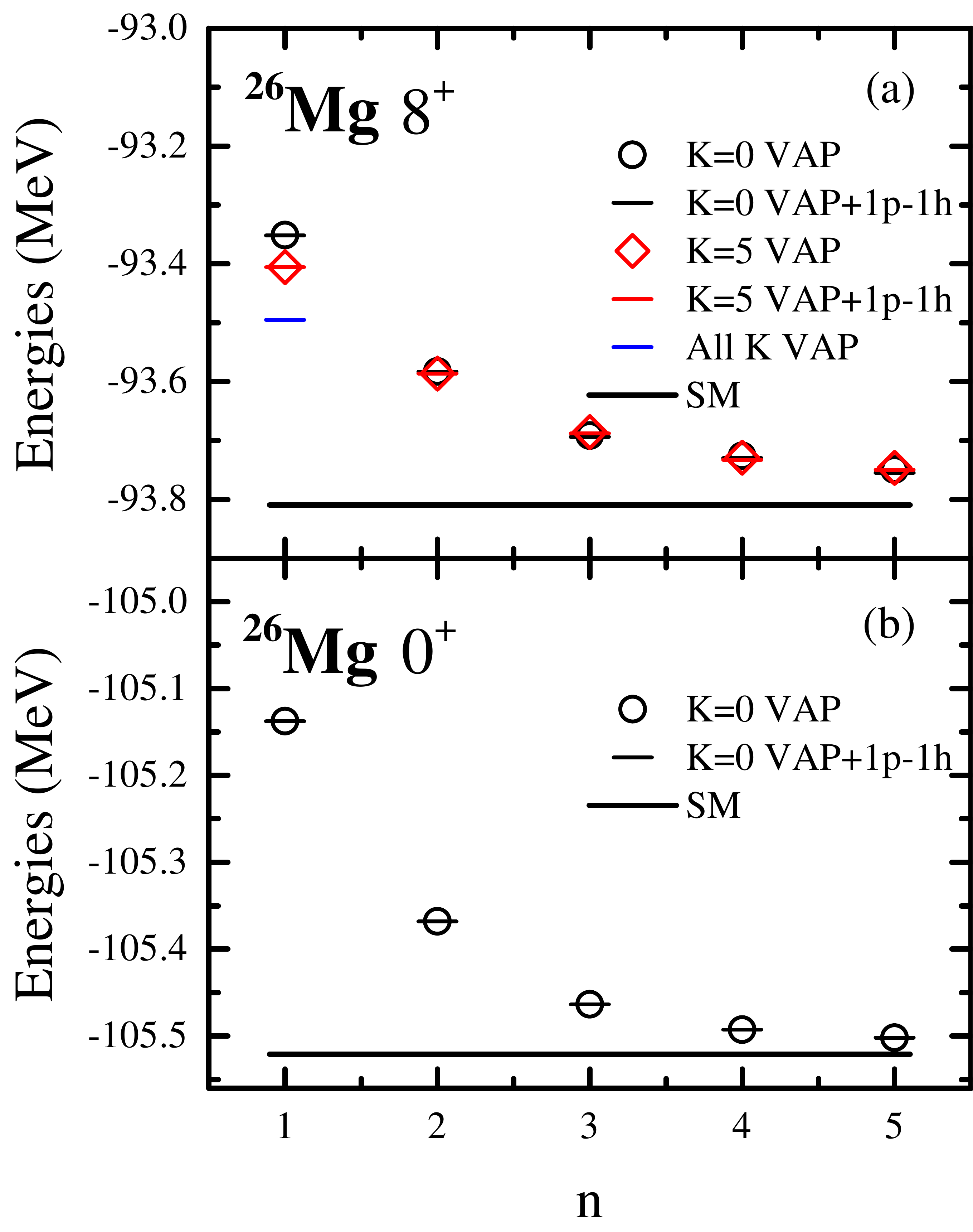}
 \caption{\label{fig2}(Color online) The converged $E^{J+}_1$ energies (black circles for $K=0$ and red diamond for $K=5$) and the corresponding $E'^{J+}_1$ energies (blak bars for $K=0$ and red bars for $K=5$ ) as functions of the number of reference SDs, $n$, for the yrast states with (a) $J^\pi$=8$^{+}$ and (b) $J^\pi$=0$^{+}$ in $^{26}$Mg. The horizonal lines show the full SM eneriges. The blue bar shows the converged energy of Eq. (\ref{wf}) taken from Fig. \ref{fig1}.
}
\end{figure}

When more reference SDs are involved, as shown in Eq. (\ref{wfks}), the situation becomes more complicated. Since the included SDs $|\Phi_{i} \rangle$ and $|\Phi_{i'} \rangle$ are independent, and the associated matrix elements
$\langle\Phi_{i} |(\hat{H}-{E'}^{J\pi}_\alpha) P_{K K}^{J\pi} A_{i'\kappa' }^{\dagger} |\Phi_{i'} \rangle$ in Eq. (\ref{hwexp}) could be arbitrary. One may expect the 1p-1h mixing may take some effect.
To  check such effect,  the $E_1^{J\pi}$ energy and its corresponding ${E'}_1^{J\pi}$ have been calculated with $n>1$.
The converged $E_1^{J\pi}$ energy and its corresponding ${E'}_1^{J\pi}$ as functions of $n$ are shown in Fig. \ref{fig2}. For the $8^+$ state in Fig. \ref{fig2}(a), the calculations are performed with $K=0$ and $K=5$, respectively, to show the $K$ independence. It is clearly seen that the $E_1^{J\pi}$ energies with $K=0$ and $K=5$ are almost coincide and they become closer and closer to the corresponding shell model energy, $e_1^{J\pi}$. However, it is striking that all the energy differences $\delta E^{J\pi}_1$ are still numerically zero for all calculated $n$ numbers.
This implies that there may exist a theorem for this interesting phenomenon.
According to the above calculations, it seems that, the situation with $E_1^{J\pi}={E'}_1^{J\pi}$ only appears after the VAP iteration converges, while the gradient of the energy becomes zero.

Let us try to dig out this theorem. For convenience, we rewrite the VAP wave function of Eq. (\ref{wfks}) for the yrast state in a simpler form,
\begin{equation}\label{swv}
|\Psi_{J\pi M1}(K)\rangle=\sum_{i=1}^{n} f_{i} P_{M K}^{J\pi}|\Phi_{i}\rangle.
\end{equation}
The corresponding energy, $E$, can be written as,
\begin{equation}\label{1p-1h5}
\begin{aligned}
E=\frac{\sum_{ii'=1}^nf_{i }^{*}{f}_{i'}\langle\Phi_i|  \hat{H}P^{J\pi}_{KK}|\Phi_{i'}\rangle}{\sum_{ii'=1}^nf_{i }^{*}{f}_{i'}\langle\Phi_i|P^{J\pi}_{KK}|\Phi_{i'}\rangle}.
   \end{aligned}
\end{equation}
Clearly, $E$ is a function of all involved variational parameters $d_{j\kappa}$, which are complex numbers,
\begin{equation}\label{1p-1h6}
d_{j\kappa}=x_{j\kappa}+i y_{j\kappa},
\end{equation}
where $x_{j\kappa}$ and $y_{j\kappa}$ are real numbers. Notice that the $d_{j}$ matrix only determines the $j$-th reference SD state $|\Phi_{j}\rangle$.
 Similar to the deductions in Ref. \cite{VAP17}, the partial derivatives $\frac{\partial E}{\partial x_{j\kappa}}$ and $\frac{\partial E}{\partial y_{j\kappa}}$ at $\vec{d}=0$ can be expressed as
\begin{eqnarray}\label{px}
\left.\frac{\partial E}{\partial x_{j\kappa}}\right|_{\vec{d}=0} &=& \sum_{i=1}^{n}f_{j }^*\langle\Phi_j|A_{j\kappa}(\hat H-E)P^{J\pi}_{KK}|\Phi_i\rangle f_i \nonumber\\ &+&\sum_{i=1}^{n}f_{i }^*\langle\Phi_i|(\hat{H}-E)P^{J\pi}_{KK}A^\dagger_{j\kappa}|\Phi_j\rangle f_j,\\
\label{py}\left.\frac{\partial E}{\partial y_{j\kappa}}\right|_{\vec{d}=0} &=&-i\sum_{i=1}^{n}f_{j }^*\langle\Phi_j|A_{j\kappa}(\hat H-E)P^{J\pi}_{KK}|\Phi_i\rangle f_i \nonumber\\ &+&i\sum_{i=1}^{n}f_{i }^*\langle\Phi_i|(\hat{H}-E)P^{J\pi}_{KK}A^\dagger_{j\kappa}|\Phi_j\rangle f_j.
\end{eqnarray}

Assuming the energy $E$ has reached a minimum at $\vec{d}=0$,  the partial derivatives in Eqs. (\ref{px}) and (\ref{py}) should be zero. Then one can get
\begin{eqnarray}\label{g1}
\sum_{i=1}^{n}f_{i }^*\langle\Phi_i|(\hat{H}-E)P^{J\pi}_{KK}A^\dagger_{j\kappa}|\Phi_j\rangle=0,
\end{eqnarray}
and its Hermitian conjugate,
\begin{eqnarray}\label{g2}
\sum_{i=1}^{n}\langle\Phi_j|A_{j\kappa}(\hat H-E)P^{J\pi}_{KK}|\Phi_i\rangle f_i=0.
\end{eqnarray}

Now, let us come back to the 1p-1h mixing. For a 1p-1h projected state,
\begin{equation}
|\Psi_{j\kappa}\rangle=P^{J\pi}_{KK} A_{j\kappa}^\dagger|\Phi_j\rangle,
\end{equation}
it may not be orthogonal to the converged VAP wave function $|\Psi_{J\pi M}(K)\rangle$ in Eq. (\ref{swv}). So one may need to get a new state by performing the Gram-Schmid orthogonalization,
\begin{eqnarray}
|\Psi'_{j\kappa}\rangle=|\Psi_{j\kappa}\rangle
-\langle \Psi_{J\pi M1}(K)|\Psi_{j\kappa}\rangle|\Psi_{J\pi M1}(K)\rangle,
\end{eqnarray}
so that
\begin{eqnarray}
\langle \Psi_{J\pi M1}(K)|\Psi'_{j\kappa}\rangle=0.
\end{eqnarray}

If there is no mixing between $|\Psi'_{j\kappa}\rangle$ and $|\Psi_{J\pi M1}(K)\rangle$, then one should have
\begin{eqnarray}
\langle \Psi_{J\pi M1}(K)|\hat H|\Psi'_{j\kappa}\rangle=0.
\end{eqnarray}
Therefore, we have
\begin{eqnarray}
\langle \Psi_{J\pi M1}(K)|(\hat H-E)|\Psi_{j\kappa}\rangle=0,
\end{eqnarray}
which is exactly the same as the Eq. (\ref{g1}). Thus, we have analytically proved that 1p-1h mixing for the yrast state does not improve the converged VAP wave function even with more reference SDs.

\section{1p-1h mixing with non-yrast states}\label{nonyrast}
As a further exploration, let us study the 1p-1h mixing with the non-yrast states. Here, the sum of the lowest energies, $S_m (m>1)$, with the same quantum numbers is minimized, as proposed in Ref. \cite{VAP18}. When the gradient of $S_m$ becomes zero, the VAP calculation converges. However, the gradient of $S_m$ being zero does not guarantee that the gradients of its members, $E^{J\pi}_\alpha$, are zero. Thus, it is expected the 1p-1h mixing might play some role in lowering the VAP energies and improving the VAP wave functions in this case.

\begin{figure}
 \centering
 \includegraphics[width=8cm]{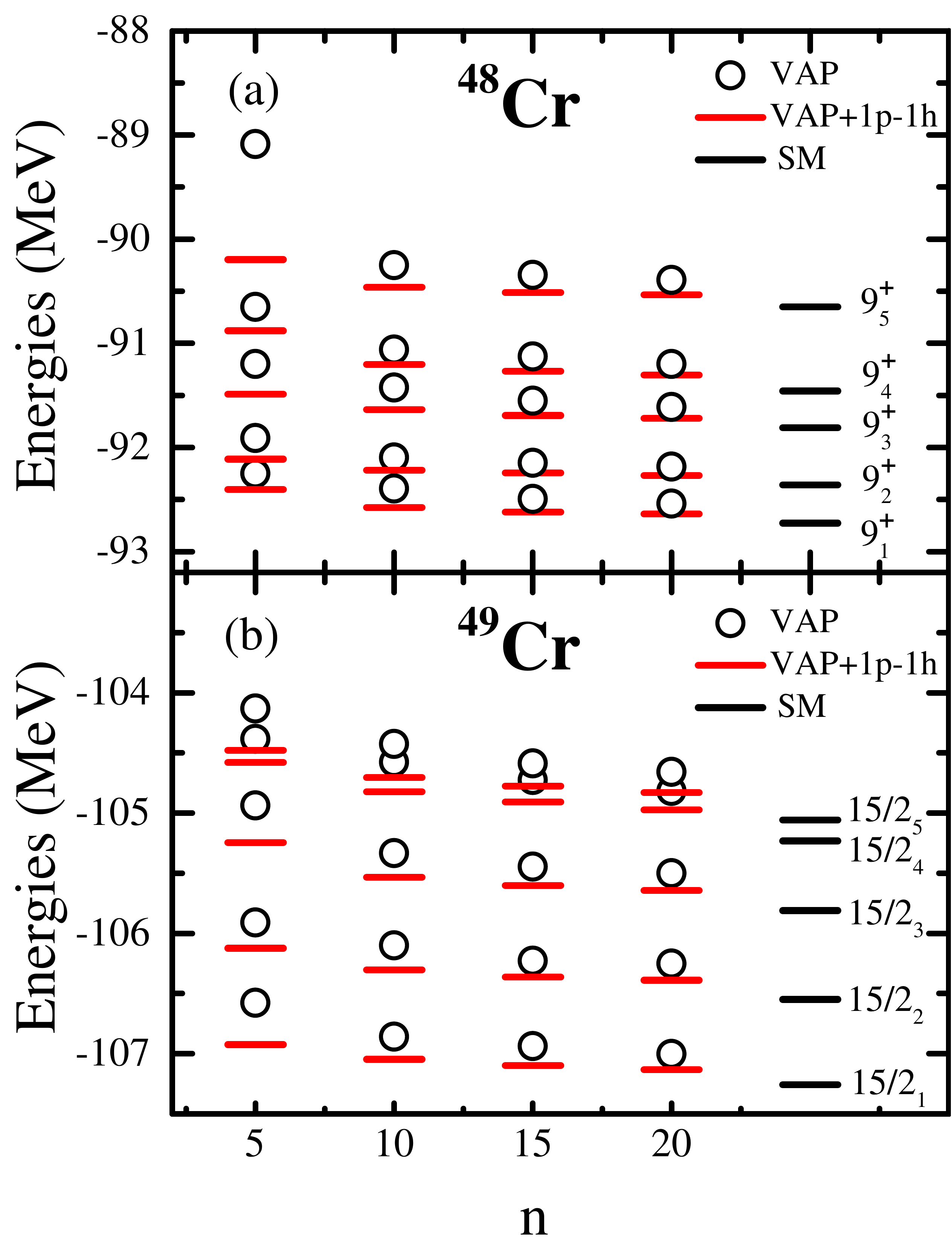}
 \caption{\label{fig3}(Color online) The converged VAP energies, $E^{J\pi}_\alpha$, (black circles)  and their corresponding $E'^{J\pi}_\alpha$ ones (red bars) calculated with different $n$ numbers for the lowest five states with (a)$J^\pi= 9^+$ in $^{48}$Cr and (b) $J^\pi=\frac{15}{2}^{-}$ in $^{49}$Cr. The SM energies $e^{J\pi}_\alpha$ are shown as black bars. The GXPF1A interaction is adopted.
}
\end{figure}

To check if such 1p-1h mixing is valid or not, some $fp$ shell nuclei are calculated using the GXPF1A interaction \cite{gxpf1a}. First of all, the sum of the lowest five $J^\pi=9^+$ energies in $^{48}$Cr and the sum of the lowest five $J^\pi=\frac{15}{2}^-$ energies in $^{49}$Cr are minimized, respectively. The trial wave function in Eq. (\ref{wfk}) is taken. The $K$ values are chosen to be $0$ for $^{48}$Cr and $1/2$ for $^{49}$Cr. For the number of selected reference SDs, we chose $n=5,10,15$ and $20$, respectively, so that one can see the changes of $E_\alpha^{J\pi}$ and ${E'}_\alpha^{J\pi}$ as functions of $n$. The calculated results are shown in Fig. \ref{fig3}. As expected, all the $E_\alpha^{J\pi}$ energies gradually drop toward the corresponding shell model ones as $n$ increases. However, different from the results in Fig. \ref{fig2}, all the calculated ${E'}_\alpha^{J\pi}$ energies are lower than the corresponding $E_\alpha^{J\pi}$ values. This means the 1p-1h mixing takes effect and indeed improves the VAP wave functions due to the nonzero gradients of the calculated VAP energies, ${E}_\alpha^{J\pi}$.

\begin{figure}
 \centering
 \includegraphics[width=8cm]{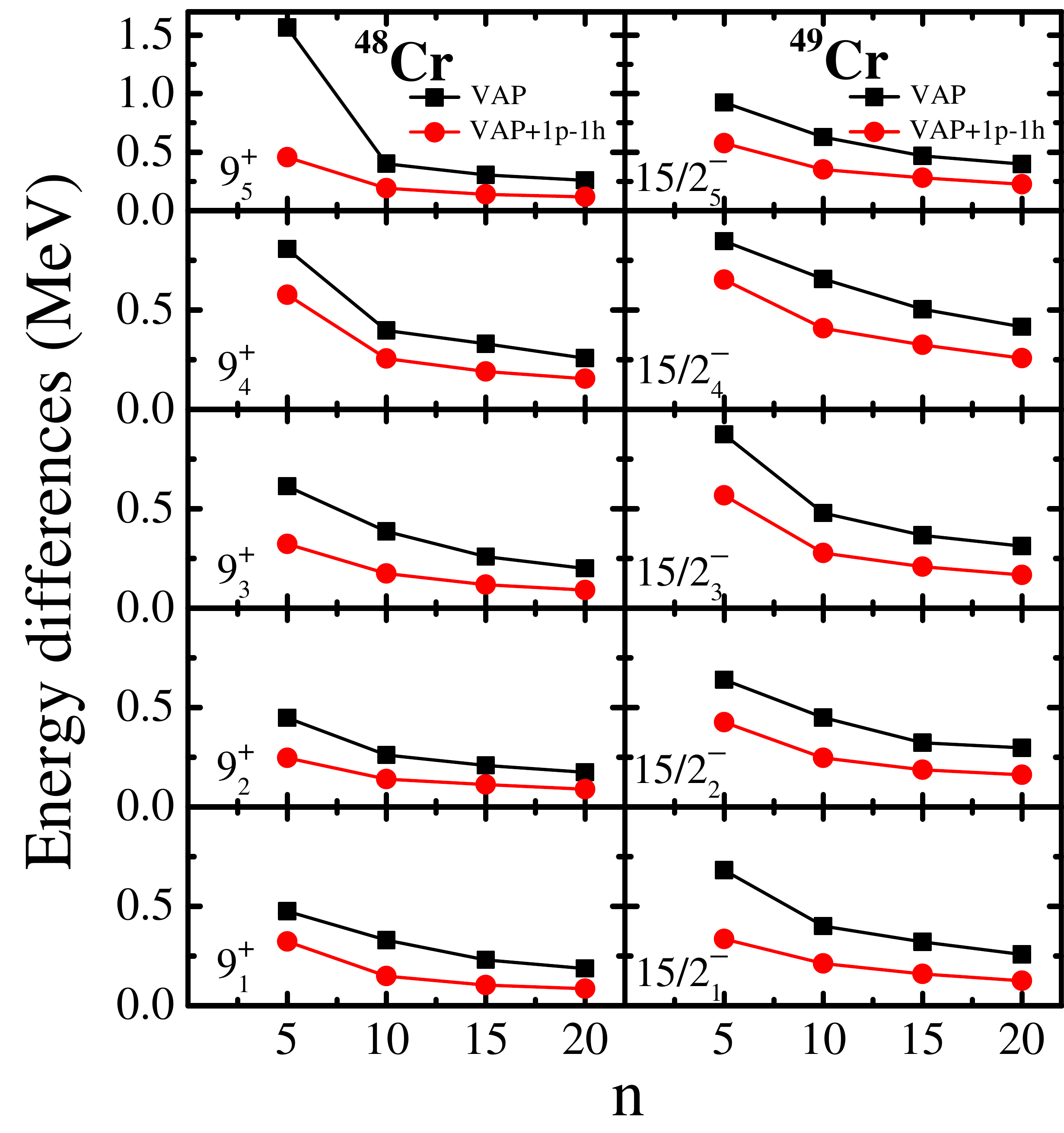}
 \caption{\label{fig4}(Color online) The energy differences, $E_\alpha^{J\pi}-e_\alpha^{J\pi}$ (black squares) and ${E'}_\alpha^{J\pi}-e_\alpha^{J\pi}$ (red dots), with different $n$ numbers. The values of $E_\alpha^{J\pi}$, ${E'}_\alpha^{J\pi}$ and $e_\alpha^{J\pi}$ are shown in Fig. \ref{fig3}.
}
\end{figure}

 The energy differences, $E_\alpha^{J\pi}-e_\alpha^{J\pi}$ and ${E'}_\alpha^{J\pi}-e_\alpha^{J\pi}$ as functions of $n$ are plotted in Fig. \ref{fig4}. Obviously, the values of ${E'}_\alpha^{J\pi}-e_\alpha^{J\pi}$ are considerably smaller than the ones of $E_\alpha^{J\pi}-e_\alpha^{J\pi}$. All the ratios of ${{E'}_\alpha^{J\pi}-e_\alpha^{J\pi}}$ to ${E_\alpha^{J\pi}-e_\alpha^{J\pi}}$ are blow 0.8. Actually, most of them are less than 0.6. This clearly shows the 1p-1h mixing can play an important role in improving a group of nuclear wave functions with the same $J^\pi$ who are varied simultaneously.

Moreover, for all calculated states in Fig. \ref{fig3}, our results show that all the energies with 1p-1h mixing, ${E'}_\alpha^{J\pi}$, at $n=10$ are even lower than the VAP energies, $E_\alpha^{J\pi}$, at $n=20$. We should mention that the computational time for the VAP calculation with $n=20$ is about 3 times longer than the one with $n=10$. Since all the matrix elements in Eq. (\ref{hwexp}) have already been evaluated during the VAP calculation, the only time spent on the 1p-1h mixing is solving the expanded HW equations of Eq. (\ref{hwexp}). Thus the 1p-1h mixing may not only improve the approximation of the VAP wave functions but also substantially save the computational time. However, the price is that the number of the included projected basis states is much larger.

\begin{figure}
 \centering
 \includegraphics[width=8cm]{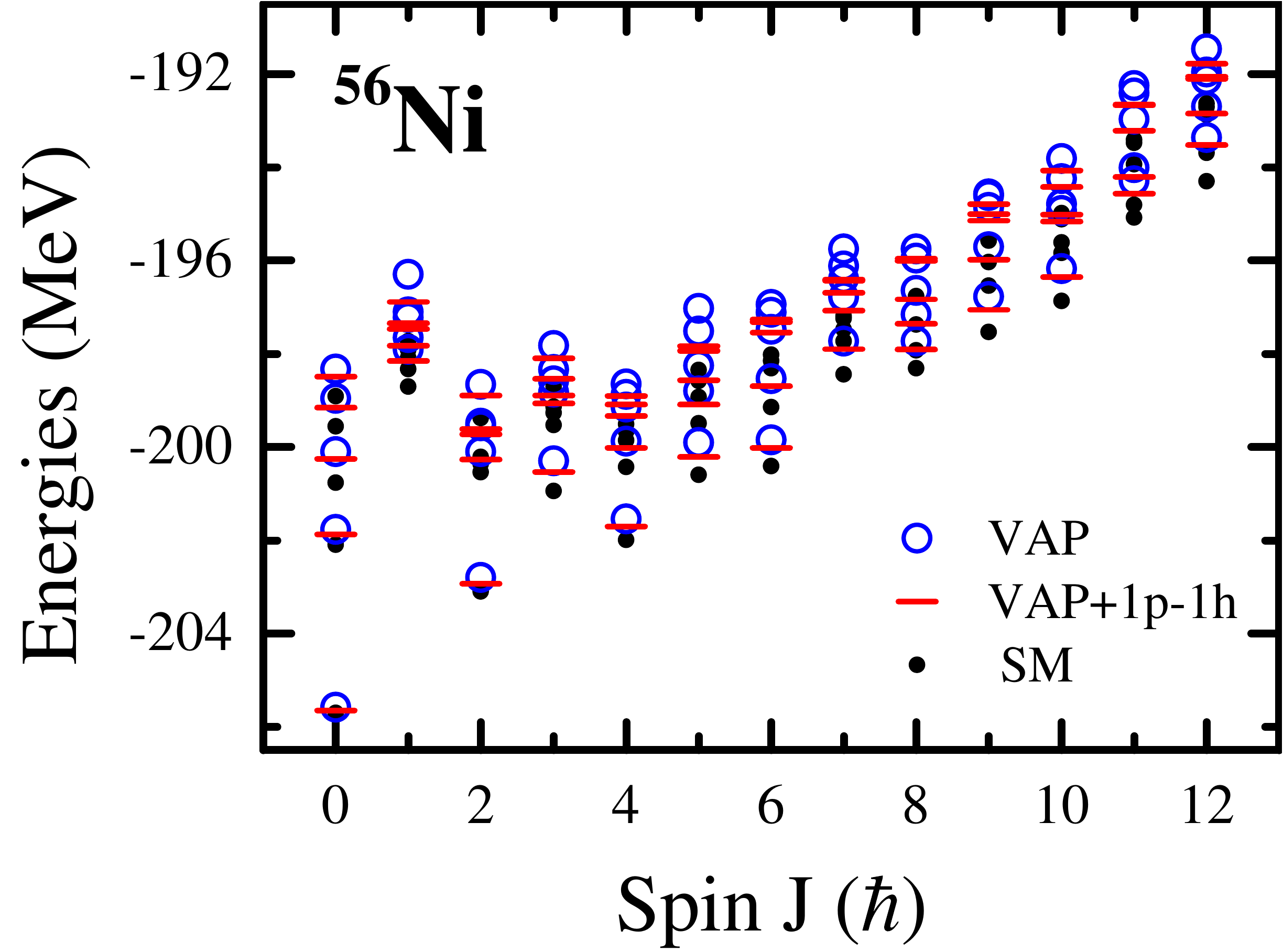}
 \caption{\label{fig5}(Color online) The lowest five VAP energies, $E_\alpha^{J\pi}$ (blue circles) and the corresponding ${E'}_\alpha^{J\pi}$ (red bars) in $^{56}$Ni with $J^\pi$ running from $0^+$ to $12^+$. The lowest five VAP energies are obtained by minimizing the sum of them at each $J^\pi$. The SM energies (black dots) are shown for comparison.
}
\end{figure}

 Finally, the VAP plus 1p-1h mixing calculations for the states with spin ranging from $J=0$ to $12$ in $^{56}$Ni are performed. For each spin, we still minimize the sum of the lowest five energies with $n=10$. The GXPF1A interaction is again adopted. The calculated results are shown in Fig. \ref{fig5}. It is seen that
 with only 10 projected SDs for each spin, the VAP method is still able to reproduce the schematic pattern of the low-lying energy levels of the full shell model. The 1p-1h mixing again makes more or less improvements for the calculated states. One may see that for some states far from the yrast line, such improvements seem more apparent.

\section{Summary and outlook}\label{sum}
 Particle-hole mixing is a natural way of improving the quality of the nuclear wave functions, as has been adopted in the random-phase approximation (RPA) \cite{PR80}. However, the 2p-2h mixing involves too large matrices that is not our present interest. In the present work, we only consider the possible 1p-1h mixing with the VAP wave functions. It is well known that in the Hartree-Fock method, the 1p-1h excited states can not be mixed into the converged HF vacuum state. Very similarly, in our VAP calculation, when only the lowest energy with a given spin and parity is minimized, we recognized that the 1p-1h projected states on top of the projected reference SDs can not be mixed into the converged VAP wave function. This interesting phenomenon has been analytically proved in the present work and can be considered as a natural generalization from the Hartree-Fock method. The common point is that for both HF and VAP, the gradients of their converged energies are zero. However, after one minimizes the sum of a group of lowest energies with the same quantum numbers in VAP, the later 1p-1h mixing may indeed give the energies lower than the corresponding VAP ones. This is because the final gradient of each energy is not guaranteed to be zero. With these results, one can easily imagine such 1p-1h mixing is indeed somewhat limited, especially for those states which energy gradients are relatively small. Of course, one may expect that the approximation of the present work could be further improved by including a number of the most important 2p-2h projected basis states. Such work will be done in the future.

On the other side, it is well known that the pairing correlation is important in the nuclear system. Such correlation can be incorporated by introducing the HFB transformation. However, in the present method, we adopt the projected HF SDs rather than the projected HFB vacuum states in order to save the computational cost. Nevertheless, it is interesting that the obtained VAP energies in the present examples are still very close to the exact shell model ones even without the explicit HFB transformation. This might because the adopted model space is rather small. However, in large model spaces, the BCS correlation may be more prominent. This could make the approximation of our method not as good as the present examples. One may think the mixing of the 1p-1h projected states not enough since the BCS correlation is non-perturbative. In this situation, we consider three possible ways to improve the approximation of the VAP wave functions. The first way is further considering the mixing of the most important $n$p-$n$h projected SDs on top of the VAP wave functions, which is very similar to our previous work \cite{PCI10}. The second way is to add more reference SDs (i.e. $n$ in Eq. (\ref{wfks}) is large) into the VAP wave function, this certainly requires more computational cost. The third way is directly taking the projected HFB vacuum states such as the VAMPIR \cite{VAMPIR04} or the QVSM \cite{Shimizu21}, but this definitely increases the computational time at least by two orders of magnitude due to the extra projections for the numbers of both proton and neutron. It would be interesting to compare these different methods in heavy deformed nuclei if all such calculations can be implemented in large model spaces in the future.

\begin{acknowledgments}
This work is supported by the National Natural Science Foundation of China under Grant Nos. $11975314$,11575290,
by the Key Laboratory of Nuclear Data foundation (JCKY2022201C158) and by the Continuous Basic Scientific Research Project Nos. WDJC-$2019$-$13$, BJ$20002501$.
\end{acknowledgments}

\appendix
\section{Basic matrix elements for the 1p-1h mixing}
Matrix elements necessary for the 1p-1h mixing are presented as a complement of those provided in Ref. \cite{lian22}, so that all basic matrix elements required in VAP calculations are available. Following the notations in Ref. \cite{lian22}, let's assume $|\Phi_a\rangle$ and $|\Phi_b\rangle$ as Slater determinants of an $N-$particle system, i.e.,
\begin{eqnarray}\label{phi}
|\Phi_a\rangle&=&\hat a^\dagger_1\cdots\hat a^\dagger_{N}|\rangle,\\
|\Phi_b\rangle&=&\hat b^\dagger_1\cdots\hat b^\dagger_{N}|\rangle.
\end{eqnarray}
According to the generalized Wick theorem, the overlap between $|\Phi_a\rangle$ and $|\Phi_b\rangle$ can be written as,
\begin{eqnarray}\label{detR}
\langle\Phi_a|\Phi_b\rangle=\langle|\hat{a}_N\cdots\hat{a}_1\hat b^\dagger_1\cdots\hat b^\dagger_{N}|\rangle=\det(R),
\end{eqnarray}
where $R$ is an $N\times N$ matrix with entries
\begin{eqnarray}
R_{ij}={\langle|\hat{a}_i\hat{b}^\dagger_j|\rangle}.
\end{eqnarray}

In VAP calculations, we need to calculate the following explicit matrix elements for the energy, the corresponding gradient and Hessian matrix,
\begin{eqnarray}\label{me}
&&\langle\Phi_a|\hat O|\Phi_b\rangle,\,
\langle\Phi_a|\hat O \hat{b}_{\mu}\hat{b}^\dagger_{\nu}|\Phi_b\rangle,\,
\langle\Phi_a|\hat{a}_{\nu}\hat{a}^\dagger_{\mu}\hat O|\Phi_b\rangle,\nonumber\\
&&\langle\Phi_a|\hat{a}_{\nu}\hat{a}^\dagger_{\mu}\hat O\hat{b}_{\mu'}\hat{b}^\dagger_{\nu'}|\Phi_b\rangle,\,
\langle\Phi_a|\hat O\hat{b}_{\mu}\hat{b}^\dagger_{\nu}\hat{b}_{\mu'}\hat{b}^\dagger_{\nu'} |\Phi_b\rangle,\nonumber\\
&&\langle\Phi_a|\hat{a}_{\nu'}\hat{a}^\dagger_{\mu'}\hat{a}_{\nu}\hat{a}^\dagger_{\mu}\hat O|\Phi_b\rangle,\label{mevap}
\end{eqnarray}
where,$1\leq \mu(\mu')\leq N$ and $N+1\leq\nu(\nu')\leq M$. Here, $M$ is the dimension of the model space. $\hat O$ refers to a one-body $\hat T$ or two-body operator $\hat V$, which can be written as
\begin{eqnarray}\label{o1}
\hat T&=&\sum_{\alpha\gamma} T_{\alpha\gamma}\hat{c}^\dagger_\alpha \hat{c}_\gamma,\\
\hat
V&=&\frac14\sum_{\alpha\beta\delta\gamma}V_{\alpha\beta\gamma\delta}
\hat{c}^\dagger_\alpha
\hat{c}^\dagger_\beta \hat{c}_\delta \hat{c}_\gamma.\label{o2}
\end{eqnarray}

In Ref. \cite{lian22}, we have presented the formulas for the following matrix elements:
\begin{eqnarray}\label{ev}
&&\langle\Phi_a|\hat O|\Phi_b\rangle,\,
\langle\Phi_a|\hat{a}_{\nu}\hat{a}^\dagger_{\mu}\hat O|\Phi_b\rangle,\nonumber\\
&&\langle\Phi_a|\hat{a}_{\nu'}\hat{a}^\dagger_{\mu'}\hat{a}_{\nu}\hat{a}^\dagger_{\mu}\hat O|\Phi_b\rangle,\nonumber
\end{eqnarray}
which are necessary for the construction of the energy variance. Then $\langle\Phi_a|\hat O \hat{b}_{\mu}\hat{b}^\dagger_{\nu}|\Phi_b\rangle$ and $\langle\Phi_a|\hat O\hat{b}_{\mu}\hat{b}^\dagger_{\nu}\hat{b}_{\mu'}\hat{b}^\dagger_{\nu'}\hat |\Phi_b\rangle$ can be obtained through the following relations,
\begin{eqnarray}\label{aa}
\langle\Phi_a|\hat O \hat{b}_{\mu}\hat{b}^\dagger_{\nu}|\Phi_b\rangle&=&\langle\Phi_b|\hat{b}_{\nu}\hat{b}^\dagger_{\mu}\hat O|\Phi_a\rangle^*,\\
\langle\Phi_a|\hat O\hat{b}_{\mu}\hat{b}^\dagger_{\nu}\hat{b}_{\mu'}\hat{b}^\dagger_{\nu'}\hat |\Phi_b\rangle&=&\langle\Phi_b|\hat{b}_{\nu'}\hat{b}^\dagger_{\mu'}\hat{b}_{\nu}\hat{b}^\dagger_{\mu}\hat O|\Phi_a\rangle^*.
\end{eqnarray}
Therefore the only left matrix elements in Eq. (\ref{mevap}) are the type of $\langle\Phi_a|\hat{a}_{\nu}\hat{a}^\dagger_{\mu}\hat O\hat{b}_{\mu'}\hat{b}^\dagger_{\nu'}|\Phi_b\rangle$ which will be reused in the 1p-1h mixing in addition to those $\langle\Phi_a|\hat O|\Phi_b\rangle$,
$\langle\Phi_a|\hat O \hat{b}_{\mu}\hat{b}^\dagger_{\nu}|\Phi_b\rangle$, and
$\langle\Phi_a|\hat{a}_{\nu}\hat{a}^\dagger_{\mu}\hat O|\Phi_b\rangle$ ones.
Thus, we only present the formulas for the $\langle\Phi_a|\hat{a}_{\nu}\hat{a}^\dagger_{\mu}\hat O\hat{b}_{\mu'}\hat{b}^\dagger_{\nu'}|\Phi_b\rangle$ matrix elements.
Based on the formulation in Ref. \cite{lian22}, one can get

\begin{eqnarray}
&&\langle\Phi_{a}|\hat{a}_{\nu} \hat{a}_{\mu}^{\dagger} \hat{T} \hat{b}_{\mu'} \hat{b}_{\nu^{\prime}}^{\dagger}| \Phi_{b}\rangle \nonumber\\
&=& \mathbb{T}_{\nu \nu^{\prime}} \overline{R}\{\mu | \mu^{\prime}\}+R_{\nu \nu^{\prime}} \overline{T}\{\mu | \mu^{\prime}\}\nonumber\\
 &&-\sum_{i, j}(\mathbb{T}_{\nu j} R_{i \nu^{\prime}}+R_{\nu j} \mathbb{T}_{i \nu^{\prime}}) \overline{R}\{i \mu | j \mu^{\prime}\}\nonumber\\
 &&-\sum_{i, j} R_{\nu j} R_{i \nu^{\prime}} \overline{T}\{i \mu | j \mu^{\prime}\},
\end{eqnarray}
where $\overline{R}\{i|j\}$, $\overline{R}\{ij|kl\}$ and $\overline{R}\{ijk|lmn\}$ can be efficiently calculated using the formulas in Ref. \cite{lian22}. $\mathbb{T}_{ij}$ is defined by
\begin{equation}
\mathbb{T}_{ij}=\langle|\hat{a}_i\hat T \hat b^\dagger_j|\rangle=\sum_{\alpha\gamma}T_{\alpha\gamma}S^+_{i\alpha}S^-_{\gamma j},
\end{equation}
where, $S^+_{i\alpha}={\langle|\hat{a}_i\hat{c}^\dagger_\alpha|\rangle}$, $S^-_{\gamma j}={\langle|\hat{c}_\gamma\hat{b}^\dagger_j|\rangle}$.
$\overline{T}\{i|j\}$ and $\overline{T}\{ij|kl\}$ can be written as,
\begin{eqnarray}
\overline{T}\{i|j\}&=&\sum_{i'j'}\mathbb{T}_{i'j'}\overline{R}\{ii'|jj'\},\\
\overline{T}\{ij|kl\}&=&\sum_{i'j'}\mathbb{T}_{i'j'}\overline{R}\{iji'|klj'\}.
\end{eqnarray}

For the two-body operator, we have
\begin{eqnarray}
&&\langle\Phi_{a}|\hat{a}_{\nu} \hat{a}_{\mu}^{\dagger} \hat{V} \hat{b}_{\mu^{\prime}} \hat{b}_{\nu^{\prime}}^{\dagger}| \Phi_{b}\rangle \nonumber\\
&=& R_{\nu \nu^{\prime}}\overline{V}\{\mu | \mu^{\prime}\}-\sum_{i k} R_{\nu k} R_{i \nu^{\prime}}\overline{V}\{i \mu | k \mu^{\prime}\} \nonumber\\
&&+\sum_{i k} \mathbb{V}_{i \nu k \nu^{\prime}} \overline{R}\{i \mu | k \mu^{\prime}\} \nonumber\\
&&-\sum_{i j k l}[R_{\nu l} \mathbb{V}_{i j k \nu^{\prime}}+R_{j \nu^{\prime}} \mathbb{V}_{i \nu k l}] \overline{R}\{i j \mu | k l \mu^{\prime}\},\nonumber\\
\end{eqnarray}
with
\begin{eqnarray}
\mathbb{V}_{i j k l}&=& \sum_{\alpha \beta \gamma \delta}V_{\alpha \beta \gamma \delta} S_{i \alpha}^{+} S_{j \beta}^{+} S_{\gamma k}^{-} S_{\delta l}^{-},\\
\overline V\{i|l\}
&=&\sum_{j<k,m<n}\mathbb{V}_{jkmn}\overline{R}\{ijk|lmn\},
\label{vil}\\
\overline V\{ij|lm\}
&=&\sum_{i'<j',l'<m'}\mathbb{V}_{i'j'l'm'}\overline{R}\{iji'j'|lml'm'\}.\nonumber\\
\label{vijlm}
\end{eqnarray}

{}

%
%

\end{document}